\documentstyle[12pt]{article}
\begin{document}
\begin{center}{\Large {\bf Nonequilibrium Magnetisation Reversal by Periodic Impulsive
Fields in Ising Meanfield Dynamics}}\end{center}

\vskip 0.5cm

\begin{center}{MUKTISH ACHARYYA}\\

{\it Department of Physics, Presidency College}\\

{\it 86/1 College Street, Calcutta-700073, India}\\

{\it  muktish.acharyya@gmail.com}\end{center}

\vskip 1 cm

\noindent  We studied the {\it nonequilibrium} magnetisation 
reversal in kinetic Ising ferromagnets driven by periodic impulsive
 magnetic field in meanfield approximation. The meanfield
differential equation was solved by sixth order Runge-Kutta-Felberg method.
The periodicity and strength of the applied impulsive magnetic field play 
the key role in magnetisation reversal. We studied the minimum strength of
impulsive field required for magnetisation reversal at any 
temperature as a function of the periodicity of the impulsive field.
In the high temperature and small period this is observed to be linear.
The results are compared with that obtained from Monte Carlo simulation
of a three dimensional Ising ferromagnet.

\vskip 1.5cm

\noindent {\bf Keywords: Ising model, Meanfield theory, 
Magnetisation reversal} 

\noindent {\bf PACS Nos:} 75.60.Jk
\newpage

\noindent {\bf I. Introduction:}

The Ising model is a prototype to study some equilibrium as well as nonequilibrium
phenomena \cite{rev}. Recently, the nonequilibrium aspects of Ising ferromagnets
in presence of time varying fields become an interesting field of modern research
\cite{rev}. The time dependent meanfield equation
\cite{tom}, of kinetic Ising ferromagnet in
presence of applied magnetic field, is a simple representation of nonequilibrium
phenomena. The nonequilibrium dynamic phase transition \cite{rev} is studied
extensively in this system. It exhibits tricritical behaviour and
a method of finding the tricritical points is proposed recently \cite{bho}. 
Like so called equilibrium transition, the nonequilibrium dynamic transition
shows the divergence of time scale and power law divergence was found recently
\cite{csd}.

The magnetisation reversal, 
{\it by nucleation},
in the Ising ferromagnet was studied by Monte Carlo
simulation (for large systems) in the presence of constant (in time) magnetic
field \cite{nucl}. This magnetisation reversal is an equilibrium phenomenon.
Due to the presence of thermal fluctuation, any small amount of field 
(constant in time) may lead to
the magnetisation reversal. In that sense no such minimum amount of field is
required. However, in the meanfield approximation a minimum amount of field
is required for magnetisation reversal. 
The externally applied periodic (in time) impulsive fields
keeps the system always away from equilibrium.
So, it would be interesting to know how the {\it nonequilibrium} 
magnetisation reversal
occurs if one applies a periodic impulsive fields. In this paper, we addressed
this question and studied the {\it nonequilibrium} magnetisation reversal 
(in kinetic Ising ferromagnets in meanfield approximation)
in presence of periodic (in time)
impulsive magnetic fields. The nonequilibrium magnetisation reversal
was studied \cite{arko}  recently in Ising model both by Monte Carlo 
simulation
and by solving the dynamical meanfield equation.
The dynamics of magnetisation reversal was also studied \cite{expt}
 experimentally
in polycrystalline Co film.

We have arranged the paper as follows: the meanfield differential equation of
kinetic Ising ferromagnet and the method of numerical solution are described
in the next section. The numerical results are given in section-III. The paper
end with a summary in section-IV.

\vskip 1cm

\noindent {\bf II. Model and numerical solution:}

The differential equation of instantaneous 
average magnetisation $m(t)$ of kinetic Ising ferromagnet, driven by 
a time varying magnetic field, in meanfield approximation, is given as \cite{tom}

\begin{equation}
\tau {{dm} \over {dt}} = -m + {\rm tanh}({{m+h(t)} \over {T}}),
\end{equation}

\noindent where, $h(t)$ is the externally applied time varying
magnetic field and $T$ is the temperature measured
in units of the Boltzmann constant ($K_B$). This equation describes the nonequilibrium
behaviour of instantaneous value of magnetisation $m(t)$ of Ising ferromagnet in 
meanfield approximation. Here, $\tau$ stands for the microscopic relaxation time for
the spin flip \cite{tom}. 
 
The periodic impulsive magnetic field is given as

\begin{eqnarray}
h(t)&=&-h_0,{\rm for~t=n\Delta t, where~n=1,2,3.}\nonumber\\
&=&0 {\rm ~~~~otherwise.~~} 
\end{eqnarray}

\noindent Where $\Delta t$ and $h_0$ denote the periodicity and strength of the
impulsive field respectively.

We have solved this
equation by sixth order Runge-Kutta-Felberg (RKF) \cite{num} method 
to get the instantaneous value
of magnetisation $m(t)$ at any finite temperature $T$, $h_0$ and $\Delta t$.
The method of solving the ordinary differential equation 
${{dm} \over {dt}} = F(t,m(t))$, by sixth order RKF method, is
described briefly as:
\begin{flushleft}{$m(t+dt) = m(t) + \left({{16k_1} \over {135}}+{{6656k_3} \over {12825}}
+{{28561k_4} \over {56430}}-{{9k_5} \over {50}}
+{{2k_6} \over {55}}\right)$}\\
{\rm where}\\
{$k_1 = dt \cdot F(t,m(t))$}\\
{$k_2 = dt \cdot F(t+{{dt} \over 4}, m+{{k_1} \over 4})$}\\
{$k_3 = dt \cdot F(t+{{3dt} \over 8}, m+{{3k_1} \over {32}}+{{9k_2} \over 32})$}\\
{$k_4 = dt \cdot F(t+{{12dt} \over {13}}, m+{{1932k_1} \over {2197}}
-{{7200k_2} \over {2197}}+{{7296k_3} \over 2197})$}
{$k_5 = dt \cdot F(t+dt,m+{{439k_1} \over 216}-8k_2+{{3680k_3} \over 513}
-{{845k_4} \over 4104})$}\\
{$k_6 = dt \cdot F(t+{{dt} \over 2}, m-{{8k_1} \over 27}+2k_2-{{3544k_3} \over 2565}
+{{1859k_4} \over 4104}-{{11k_5} \over 40})$.................................(3)}
\end{flushleft}

\noindent The time interval $dt$ was measured in units of 
$\tau$ (the time taken to flip a single
spin). Actually, we have used $dt = 0.01$ (setting $\tau$=1.0). 
The local error involved \cite{num} in the sixth order RKF method
is of the order of $(dt)^6 (=10^{-12})$. 
We started with initial condition $m(t=0) = 1.0$.

\vskip 1cm

\noindent {\bf III. Results:} 

Starting from the initial magnetisation $m(t=0)=1.0$, the instantaneous 
magnetisation ($m(t)$) decreases as time goes on. At any instant the
magnetisation $m(t)$ becomes negative and this is known as magnetisation reversal.
We studied this magnetisation reversal by applying the periodic impulsive
field (Eqn.-2). The time required for magnetisation reversal depend on the 
value of $\Delta t$ and $h_0$ and the temperature $T$ and is called
reversal time $T_r$. 
All times are measured in the unit of $0.01\tau$ throughout this study.
A typical results of 
magnetisation reversal is shown in Fig.-1, for $\Delta t = 10$ 
and $h_0 = 1.5$ at temperature $T=0.90$. Here, the magnetisation reversal 
occurs at time $t=760$. 
It was observed that, at any temperature $T$, the 
reversal occurs early as we increase the strength $h_0$ keeping $\Delta t$ fixed.
In this case, a minimum value of strength, $h^r_{min}$, is required for reversal.
We checked few cases, taking the value of $h_0$ less than $h^r_{min}$ and observed
no reversal even for $t=5\times10^6$. 

At any fixed temperature $T$, the minimum field strength ($h^r_{min}$)
required for magnetisation reversal, decreases as the periodicity 
($\Delta t$) decreases. Similarly, for fixed $\Delta t$, $h^r_{min}$
decreases as the temperature increases. In the present study, we observed
that, at temperature $T=0.80$, 
for $\Delta t = 6$ and $\Delta t = 8$ the values of $h^r_{min}$
becomes 0.33 and 0.43 respectively. 
Similarly, at temperature $T=0.90$, for $\Delta t = 6$
and $\Delta t = 8$ the values of $h^r_{min}$ become 0.13 and 0.17 respectively.
It may be noted here that for static (in time) field the 
minimum values calculated for {\it equilibrium} magnetisation reversal
are 0.07 and 0.03 for $T=0.80$ and $T=0.90$ respectively.
At any particular temperature, the value of minimum field required 
(for periodic impulsive fields in nonequilibrium
case) is always higher than that for the static field.
These results are shown in Fig.-2. In this figure, the reversal time $T_r$
is plotted against the field strength $h_0$ for different values of $\Delta t$
and temperature $T$. The reversal time $T_r$
is observed to increase here with the decrease of field strength $h_0$.

At any fixed temperature $T$, the minimum required field strength $h^r_{min}$,
increases with the periodicity $\Delta t$ of the impulsive fields. 
In the high temperature ($T_c=1$ for equilibrium ferro-para transition)
limit, it is observed
that the relationship is linear (for small value of $\Delta t$). 
We have shown this in Fig.-3.

These results are compared with that obtained from Monte Carlo simulation.
We have considered a three dimensional Ising model with
nearest neighbour ferromagnetic interaction 
under periodic boundary condition in all
directions. The Hamiltonian of this system is represented as

\begin{equation}
H= -J\Sigma_{<ij>}S_iS_j - h(t)\Sigma_i S_i
\end{equation}

\noindent where $S_i=\pm 1$ is Ising spins, $J$ is the nearest neighbour
ferromagnetic interaction strength and $h(t)$ is periodic impulsive field
(Eqn-2). We have considered a cubic lattice of size $L=20$ (here).
The initial condition was taken as all spins are up (i.e., $S_i = +1$ for
all $i$). The updating scheme is taken as random. 
We select a spin randomly
and calculated the energy in flipping, i.e., $\Delta E$. We flipped the
selected spin with probability $P_f = {\rm Min}[1,{\rm exp}
(-\Delta E/K_BT)]$, 
following the Metropolis algorithm\cite{binder}. 
Here $K_B$ is Boltzmann constant. The
temperature $T$ is measured in the unit of $J/K_B$. 
$L^3$ number of such random updating of Ising spins constitutes a single
Monte Carlo step. This defines the unit of time in this simulation.
In three dimensions
the ferro-para transition temperature is $T_c = 4.511$ \cite{binder}.
Now we allowed the system to follow the Metropolis dynamics in the
presence of periodic impulsive field $h(t)$. At any fixed temperature
$T$, we have calculated (averaged over 10 different random updating
sequences) the minimum field required $h^r_{min}$ for
the reversal of magnetisation and studied it as a function of the
interval or periodicity $\Delta t$ of periodic impulsive field. 
Here also, we have observed the linear variation of $h^r_{min}$ with
$\Delta t$. We have studied this for two different temperatures, here,
$T=4.0$ and $T=4.2$. We observed that as the temperature increases the
slope of the straight line ($h^r_{min}$ versus $\Delta t$)
decreases. This is shown in Fig-4. These Monte Carlo study 
supports the meanfield results.

\vskip 0.5cm

\noindent {\bf IV. Summary:}

We have studied the {\it nonequilibrium} magnetisation reversal
 of kinetic Ising ferromagnets
in presence of periodic impulsive magnetic fields in meanfield approximation.
At any finite temperature and fixed value of the periodicity of the impulsive
field a minimum value of field strength 
is required for magnetisation reversal (sign
change of instantaneous magnetisation). This value of minimum required field strength
depends on the periodicity of the impulsive fields. As the periodicity increases
the minimum field required increases. 
In the high temperature limit, it is observed to be linear 
in the periodicity (for small values)
of the applied impulsive fields. 
This study differs from the case where the magnetic field is constant
in time and consequently this leads to equilibrium magnetisation reversal.
However, in this case since the field is time dependent, the 
characteristics of the magnetisation reversal is nonequilibrium type.
These are studied in Ising meanfield dynamics. As a comparison the the
nonequilibrium magnetisation reversal is also studied in three dimensional 
Ising ferromagnet driven by periodic impulsive fields by Monte Carlo
simulation. The Monte Carlo results and the meanfield results are in
good agreement.

\vskip 1cm

\noindent {\bf Acknowledgments:} The author would like to thank
Urbashi Satpathi  and Abhirup Patra for 
collecting few references. He would also like to thank
Pradip Mukherjee for helping in typesetting of the manuscript.

\newpage
\setlength{\unitlength}{0.240900pt}
\ifx\plotpoint\undefined\newsavebox{\plotpoint}\fi
\sbox{\plotpoint}{\rule[-0.200pt]{0.400pt}{0.400pt}}%


\vskip 1cm

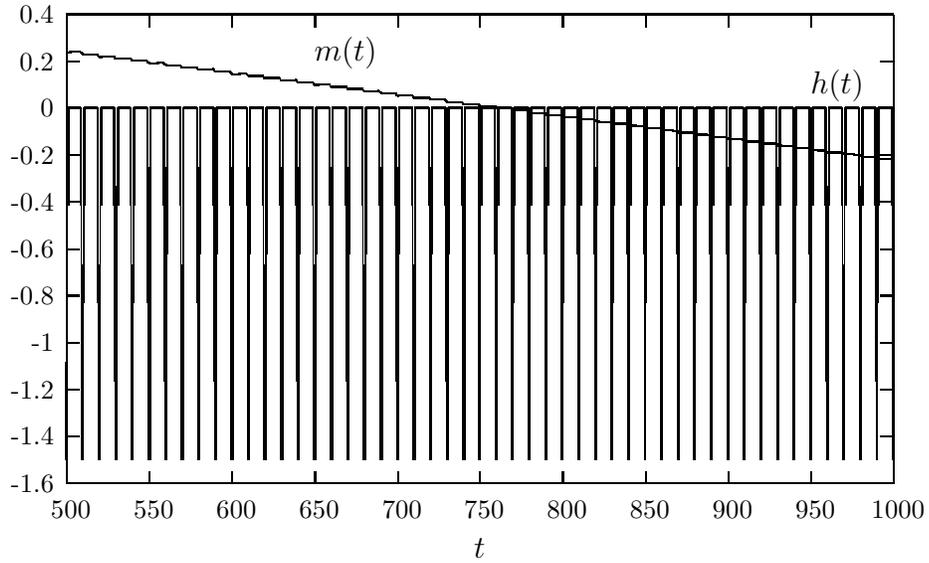
\noindent {\bf Fig-1.} The time variation of magnetisation $m(t)$ and the
applied periodic impulsive field $h(t)$ of periodicity $\Delta t = 10$ and
strength $h_0=1.5$. The magnetisation reversal happens at $t=760$. Here,
the temperature $T=0.90$.

\newpage

\setlength{\unitlength}{0.240900pt}
\ifx\plotpoint\undefined\newsavebox{\plotpoint}\fi
\sbox{\plotpoint}{\rule[-0.200pt]{0.400pt}{0.400pt}}%
%

\vskip 1cm

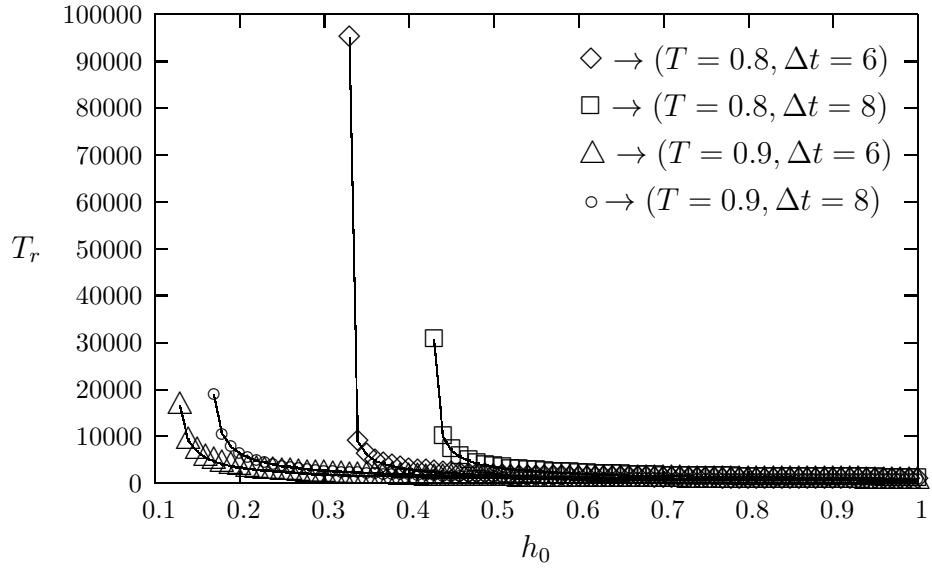
\noindent {\bf Fig-2.} The variation of magnetisation reversal time ($T_r$)
with respect to the strength of impulsive field ($h_0$) for different temperatures
($T$) and periodicity ($\Delta t$) of the impulsive fields.

\newpage

\setlength{\unitlength}{0.240900pt}
\ifx\plotpoint\undefined\newsavebox{\plotpoint}\fi
\sbox{\plotpoint}{\rule[-0.200pt]{0.400pt}{0.400pt}}%
%

\vskip 1cm

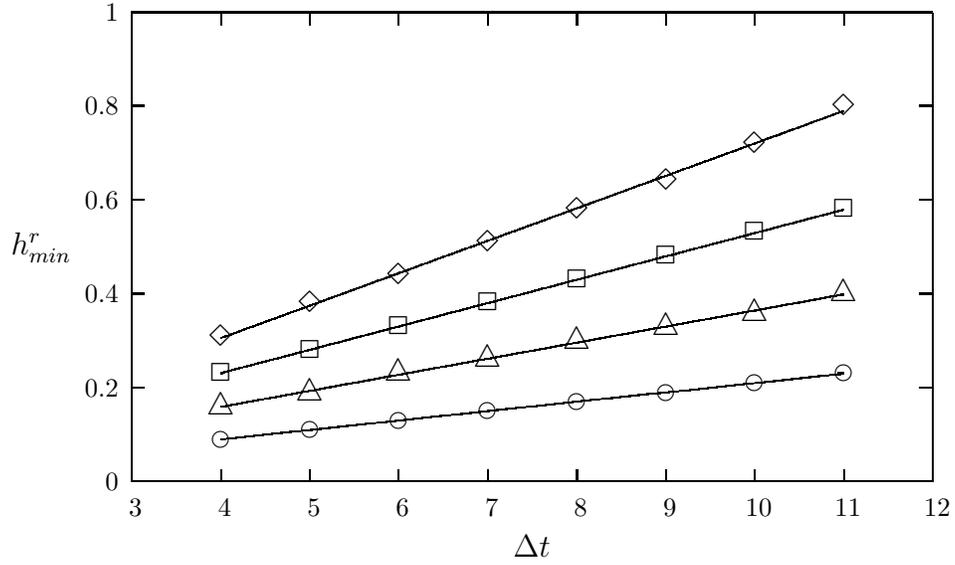
\noindent {\bf Fig-3.} The variation of $h^r_{min}$ with $\Delta t$ for different
temperatures ($T$) represented by different symbols. $T=0.75(\Diamond)$, 
$T=0.80(\Box)$, $T=0.85(\triangle)$ and $T=0.90$(O). Continuous straight lines represent
the linear best fit.

\newpage

\setlength{\unitlength}{0.240900pt}
\ifx\plotpoint\undefined\newsavebox{\plotpoint}\fi
\sbox{\plotpoint}{\rule[-0.200pt]{0.400pt}{0.400pt}}%
%

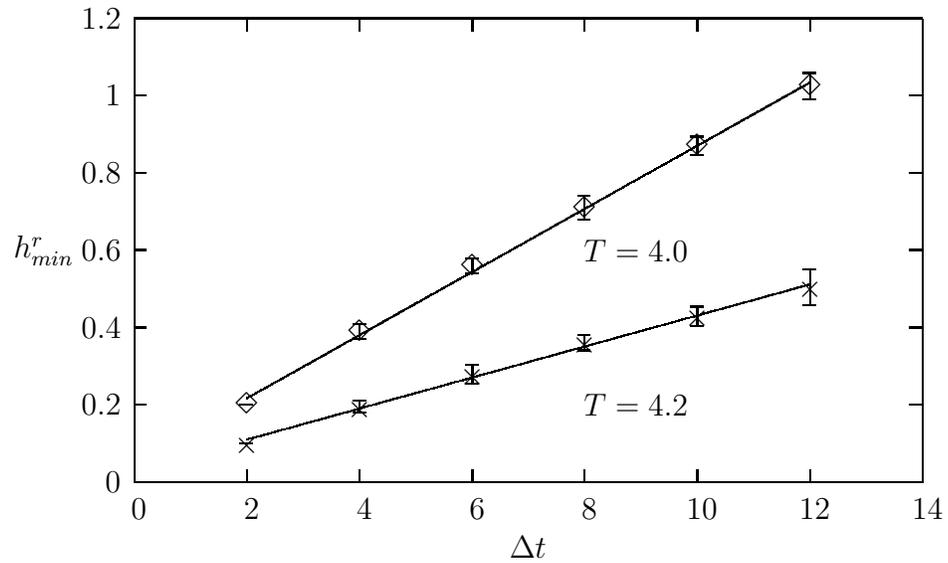
\noindent {\bf Fig.-4.} The plots of $h^r_{min}$ versus $\Delta t$ for
two different temperatures, obtained from Monte Carlo simulation. 
The errorbars in each data are shown by
small vertical lines. The continuous straight line represents
the linear best fit.
\end{document}